\documentclass[sigconf]{acmart}

\AtBeginDocument{%
  }

\newcommand{\MethodName} {Opti3DGS}

\usepackage{xcolor}
\usepackage{colortbl}
\usepackage{multirow}

\copyrightyear{2025}
\acmYear{2025}
\setcopyright{acmlicensed}\acmConference[CVMP '25]{European Conference on Visual Media Production}{December 3--4, 2025}{London, United Kingdom}
\acmBooktitle{European Conference on Visual Media Production (CVMP '25), December 3--4, 2025, London, United Kingdom}
\acmDOI{10.1145/3756863.3769707}
\acmISBN{979-8-4007-2117-5/2025/12}

\begin{document}

\title{Optimized 3D Gaussian Splatting using \\Coarse-to-Fine Image Frequency Modulation}

\author{Umar Farooq}
\affiliation{%
  \institution{University of Surrey}
  \city{Guildford}
  \country{United Kingdom}
}
\email{m.farooq@surrey.ac.uk}

\author{Jean-Yves Guillemaut}
\affiliation{%
	\institution{University of Surrey}
	\city{Guildford}
	\country{United Kingdom}
}
\email{j.guillemaut@surrey.ac.uk}

\author{Graham Thomas}
\affiliation{%
	\institution{University of Surrey}
	\city{Guildford}
	\country{United Kingdom}
}
\email{g.a.thomas@surrey.ac.uk}

\author{Adrian Hilton}
\affiliation{%
	\institution{University of Surrey}
	\city{Guildford}
	\country{United Kingdom}
}
\email{a.hilton@surrey.ac.uk}

\author{Marco Volino}
\affiliation{%
	\institution{University of Surrey}
	\city{Guildford}
	\country{United Kingdom}
}
\email{m.volino@surrey.ac.uk}

\renewcommand{\shortauthors}{Farooq et al.}

\begin{abstract}
	The field of Novel View Synthesis has been revolutionized by 3D Gaussian Splatting (3DGS), which enables high-quality scene reconstruction that can be rendered in real-time.
	3DGS-based techniques typically suffer from high GPU memory and disk storage requirements which limits their practical application on consumer-grade devices.
	We propose \MethodName, a novel frequency-modulated coarse-to-fine optimization framework that aims to minimize the number of Gaussian primitives used to represent a scene, thus reducing memory and storage demands.
	\MethodName~ leverages image frequency modulation, initially enforcing a coarse scene representation and progressively refining it by modulating frequency details in the training images.
	On the baseline 3DGS \cite{kerbl20233d}, we demonstrate an average reduction of 62\% in Gaussians, a 40\% reduction in the training GPU memory requirements and a 20\% reduction in optimization time without sacrificing the visual quality.
	Furthermore, we show that our method integrates seamlessly with many 3DGS-based techniques \cite{papantonakis2024reducing,lee2023compact,girish2024eagles,fang2024mini,mallick2024taming}, consistently reducing the number of Gaussian primitives while maintaining, and often improving, visual quality.
\end{abstract}




\begin{teaserfigure}
  \includegraphics[width=\textwidth]{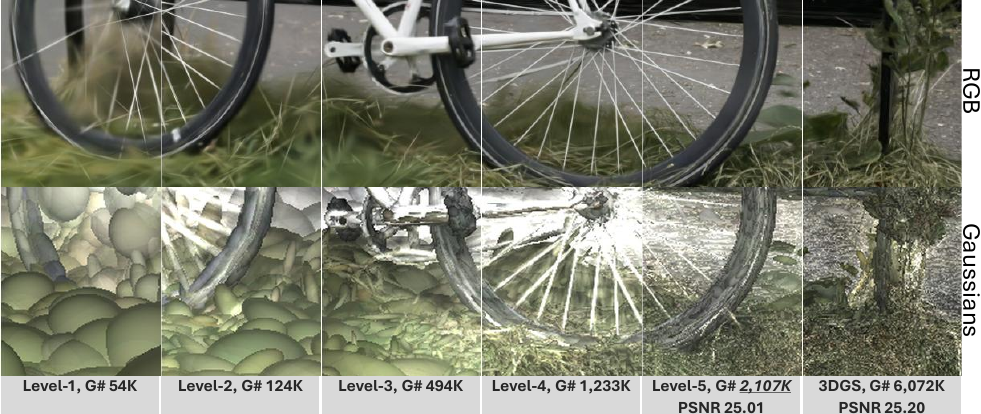}
  \caption{\MethodName~progressively grows levels to reduce redundant and unused Gaussians while maintaining accurate geometric structure and competitive quality.}
  \label{fig:teaser}
\end{teaserfigure}

\maketitle

%
% Introduction
%
\section{Introduction}

\begin{figure*}[t]
	\centering
	\includegraphics[width=\linewidth]{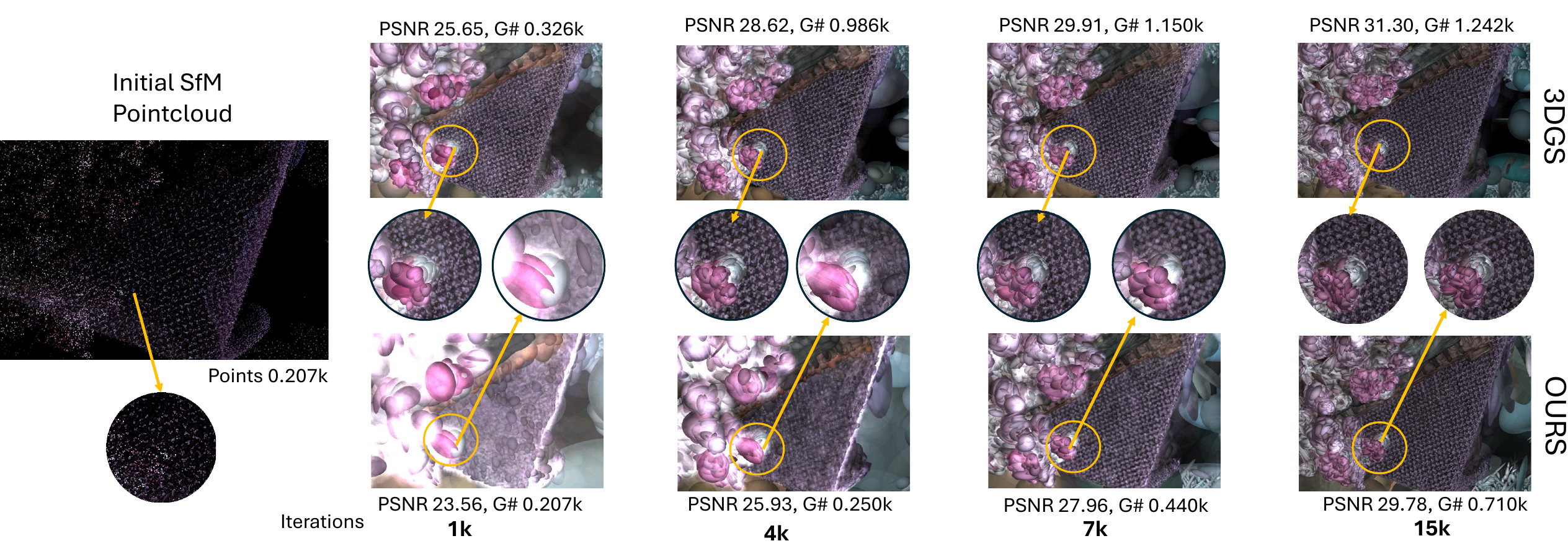}
	\caption{
		We start with uniform distribution of larger Gaussians as compared to 3DGS which starts heavy densification from the very start of optimization thus complicating the loss landscape. The zoomed views show efficient use of the initial sparse pointcloud by our approach.
	}
	\label{fig:gaussianSize}
	\vspace{-2mm}
\end{figure*}

Photorealistic scene reconstruction remains a fundamental challenge in 3D computer vision, with applications spanning augmented and virtual reality, media production, and immersive gaming.
Despite significant advancements, existing Novel View Synthesis (NVS) methods - particularly those based on Neural Radiance Fields (NeRF) \cite{mildenhall2021nerf} - face substantial computational constraints \cite{hu2022efficientnerf, li2023nerfacc, deng2023compressing}. 
These constraints hinder broader adoption, especially on web, mobile, and edge devices, where efficient rendering and low memory usage are critical.

3D Gaussian Splatting (3DGS) \cite{kerbl20233d}, recently introduced as an explicit point-based NVS technique, achieves high visual fidelity with fast training and real-time rendering. 
However, it suffers from significant GPU memory and storage overhead \cite{papantonakis2024reducing, niedermayr2023compressed}, limiting its applicability across diverse device form factors.

To address these challenges we introduce \MethodName,  a novel and efficient coarse-to-fine optimization strategy that reduces the number of Gaussians used to represent a scene. 
This translates into reduced memory and storage overheads compared to 3DGS \cite{kerbl20233d}.
Our approach begins with intentional over-reconstruction, which serves as an effective regularizer, followed by progressive refinement of the scene to balance detail and efficiency.
Several techniques have been proposed which aim to reduce 3DGS memory and storage costs.
Some focus on compressing the stored representation \cite{niedermayr2023compressed, lee2023compact, fan2023lightgaussian, navaneet2023compact3d, liu2024compgs}, others optimize training speed and parallelization \cite{zhao2024scaling}, while a few attempt to minimize Gaussian primitive count during optimization itself \cite{lee2023compact, papantonakis2024reducing}.
Our approach falls into the latter category, reducing Gaussian primitives during the optimization process, thereby decreasing GPU memory demands and accelerating training speed while maintaining visual quality. 
Notably, our method achieves these benefits as a plug-in approach without introducing additional learnable parameters or requiring lengthy optimizations.

\MethodName~improves the baseline 3DGS \cite{kerbl20233d}, achieving a 62\% reduction in Gaussian primitives, a 40\% decrease in GPU memory usage during training, and a 20\% speedup in optimization time. 
Our technique is highly modular and can be easily incorporated into most 3DGS-based pipelines with minimal modifications. 
Furthermore, we demonstrate its positive impact when integrated into optimized 3DGS variants such as MiniSplatting \cite{fang2024mini} and Taming3DGS \cite{mallick2024taming}, as well as other recent appraches \cite{papantonakis2024reducing,lee2023compact,girish2024eagles}. 
Due to the coarse-to-fine nature of our approach, it naturally produces level-of-detail representation as a byproduct, shown in Figure \ref{fig:teaser}, which could have useful applications in foveated rendering, streaming applications and advanced storage mechanisms.
To summarize, our key contributions are:
\begin{itemize}
	\item We present \MethodName, a framework to minimize the number of Gaussian primitives in a 3DGS scene resulting in improved  training speed,  memory and storage requirements.
	\item We propose a coarse-to-fine optimization strategy that increases the size of the Gaussians at the start of optimization and then slowly breaks down the large coarse Gaussians in a controlled fashion leading to an improved Gaussian densification process.
	\item We demonstrate that \MethodName~can be seamlessly integrated into a wide range of existing 3DGS techniques minimizing the number of Gaussians used to represent the scene without sacrificing visual quality, and in most cases improving it.
	
\end{itemize}

\section{Related Work}
\noindent\textbf{3DGS Compression}: One approach to reduce the storage size of 3DGS scenes focuses on quantizing raw parameters stored for each Gaussian primitive \cite{niedermayr2023compressed,lee2023compact,fan2023lightgaussian,navaneet2023compact3d,liu2024compgs}.
These enable efficient storage and transmission but since the optimization is done after 3DGS scene optimization, it does not lower GPU memory requirements during training and some of these methods require cumbersome decoding in GPU memory from a compressed representation which can be computationally expensive to render \cite{kim2024color}.
Some methods like \cite{girish2024eagles} also include an efficient encoding and decoding mechanism for parameters during the optimization process, leading to both storage and compute time gains. 
However, in the case of \cite{girish2024eagles} they introduce additional learnable parameters and use a separate neural network to decode the color values and face complications due to quantized vectors not being differentiable.
\cite{girish2024eagles} also use progressive resolution training which is different from our frequency based image filtering as outlined in section \ref{sec:results}.
A subset of these methods \cite{niedermayr2023compressed, girish2024eagles, fang2024mini, wang2024end, papantonakis2024reducing} also compute the rendering importance of each Gaussian primitive based on its contribution to individual pixels, its relative size or the Gaussian neighborhood it is located in.

\noindent\textbf{Gaussian Primitive Reduction}:
An alternative approach is to actively reduce the number of primitives during the optimization process \cite{lee2023compact, papantonakis2024reducing, fang2024mini, mallick2024taming}. 
The Adaptive Density Control (ADC) mechanism in \cite{kerbl20233d} removes Gaussians below a fixed opacity threshold at predefined intervals and densifies ones with large gradients by either splitting or cloning them. 
But despite these measures the primitive count increases rapidly and with much variation across runs\cite{mallick2024taming}.
\cite{lee2023compact} proposed to learn a volume aware mask during training, incurring additional learnable parameters and optimization time, to remove primitives which are either small or have minimal contribution. %
But this approach is limited in its effectiveness because the problem stems from dense clustering of Gaussians in certain locations with a significant number of them not even contributing to the color at all \cite{niedermayr2023compressed, fang2024mini}. 
\cite{fang2024mini, yang2025localized} show that there are multi-view inconsistencies between the centroids of the Gaussians and their location density is not uniform even in areas with uniform color and texture attributes and propose to regularize this inconsistency. 
\cite{papantonakis2024reducing} prunes Gaussians during the optimization process based on their rendering importance and signs of local clustering, however they do not reduce the peak GPU memory requirements during the optimization process.
\cite{mallick2024taming} propose the use of per-view per-pixel saliency maps to guide the densification process in a selective manner with predictable growth trend and used fused operations to speed up optimization. 
However all of these approaches require major changes in the ADC functionality and thus are not easily inter-compatible with each other and have unknown effects if combined. Whereas, \MethodName~ provides a plug-in approach that can be combined with various pipelines with almost no change to the core algorithm.
\newline
\cite{girish2024eagles} also use a coarse-to-fine strategy but the downsize images in resolution progressively, in contrast \MethodName~modulates the image frequency with the image resolution staying fixed. 
Both approaches have a distinct impact on the image pixels and quite different results. 
\cite{zhang2024fregs} also use a frequency-based approach for optimizing 3DGS but they provide gradient signals in multiple separate frequency bands simultaneously with the goal of increasing visual quality which leads to a significantly higher Gaussian count.
\newline
\noindent\textbf{Summary}: We propose \MethodName, a coarse-to-fine frequency modulation approach to reduce the number of Gaussians by only changing the training images during the optimization process. 
\MethodName~ sits behind the ADC and is not directly involved in the differential rendering process and no changes need to be made in the core algorithm.
This ensures wide ranging plug-in compatibility with most 3DGS based methods and also allows \MethodName~to integrate with compression and quantization-based methods. 
\MethodName~also directly minimizes the Gaussians which leads to reduced GPU memory requirements during the optimization process reducing overall GPU memory requirements.
\begin{figure*}[ht!]
	\centering
	\includegraphics[width=\linewidth]{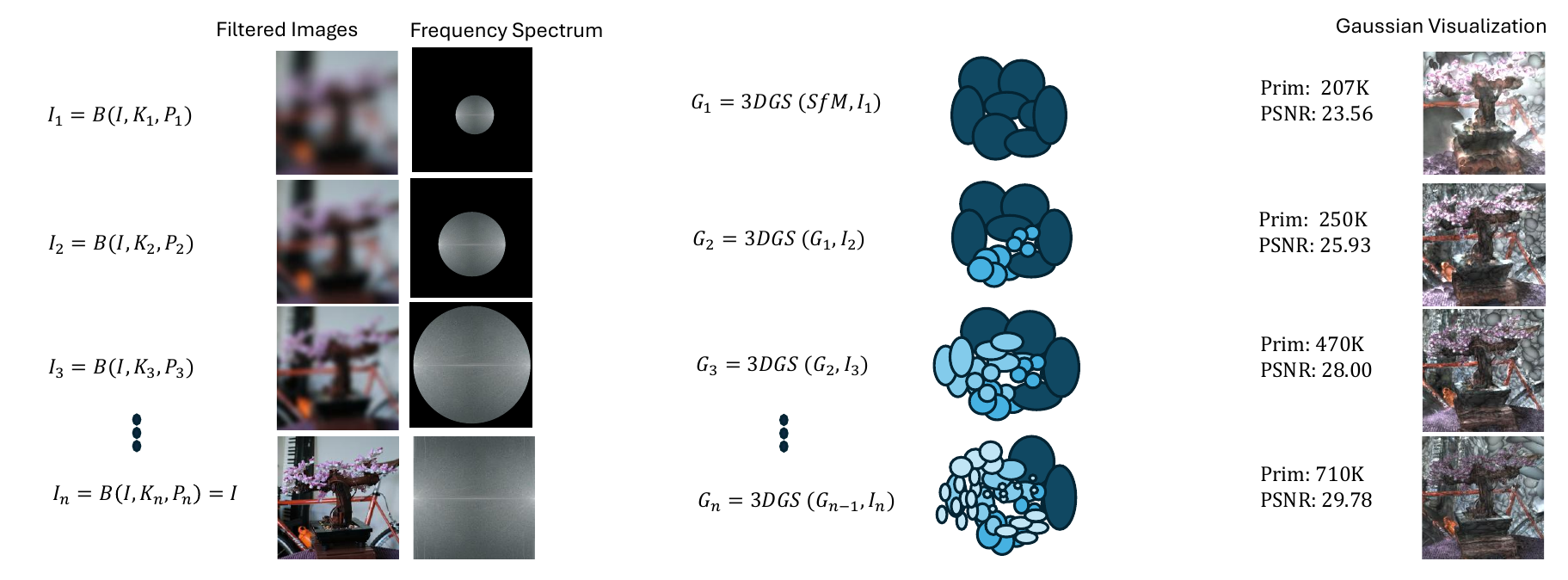}
	\caption{
        Method Overview: We start the optimization with very large and coarse Gaussians which are then densified based on the differentiable rendering signal. Specifically we control the frequency spectrum of the training images, gradually increasing the threshold for the low pass filtering.
	}
	\label{fig:methodDiagram}
\end{figure*}

\section{Method}

\subsection{Preliminaries: 3D Gaussian Splatting}
3DGS \cite{kerbl20233d} uses anisotropic Gaussians whose shape, color, opacity and refraction parameters are estimated via multi-view image-based optimization. 
Each Gaussian is defined by a covariance matrix $\Sigma$, a center position $\mathbf{p}$, opacity $\alpha$, and spherical harmonics coefficients for color $\mathbf{c}$.
The covariance matrix $\Sigma$ is decomposed into a scaling matrix and a rotation matrix to facilitate differentiable optimization.
For rendering, all Gaussians in the view frustum are splatted using onto a 2D plane followed by $\alpha$-blending of the color based on Gaussian opacity to achieve the final image.
The color $C$ of a pixel is computed by blending $N$ ordered 2D Gaussians that overlap the pixel, formulated as:
\begin{equation}
	\footnotesize
	\label{eq:base}
	C = \sum_{i \in N} c_i \alpha_i \prod_{j=1}^{i-1} (1 - \alpha_j)
\end{equation}
Here, the final color $c_i$ of a pixel is defined by the color of each Gaussians, dependent on the spherical harmonics based on the view direction, falling on to the specific pixel multiplied by its individual opacity and the cumulative opacity $\alpha_i$ of the $i$ Gaussians in front.
This approach ensures that the contributions of overlapping Gaussians are blended accurately, resulting in a precise and realistic rendered image of the scene.
The implementation also allows gradients to pass to an unrestricted number of Gaussians at any depth behind the camera position.
Over the course of optimization, numerous tiny changes based on the gradients from differential rendering of multi-view images allow for the Gaussians to approximate shapes and colors close to real scene geometry.

\subsection{\MethodName~Overview}
Our approach, shown in Figure \ref{fig:methodDiagram}, controls the frequency information in the training images which has a downstream effect on the ADC mechanism by providing it with a streamlined loss landscape.
Starting with low frequency details in the scene, it steers the densification process to first approximate the large scale structure of the scene using large and coarse Gaussians, as shown in Figure \ref{fig:teaser}.
This also alleviates the pressure on the ADC to densify the image in regions with highly dense initial Structure-from-Motion (SfM) points, as shown in Figure \ref{fig:gaussianSize}.
This indirect modulation signal allows the optimizer to approach a desirable balance between focusing on under and over-reconstructed regions of the scene.

\subsection{Image Frequency Modulation}
Digital images are made up of a spectrum of frequencies representing the structure and pattern of the color in them, with low frequencies corresponding to structure in the pixel space with gradual changes and high frequencies representing frequently changing color patterns like textures or white noise.
There are a number of techniques for manipulating, enhancing or removing signals in different parts of the frequency spectrum depending on the downstream use-cases.
Conceptually, a simple way of removing high frequency signals from an image exploits the local relation between pixels in a neighborhood to tease out only the consistent and stable color structures.
The opposite can be thought to remove low frequency signals, but in practice a wide range of mathematical and statistical machinery is employed to perform various frequency restriction operations on images in an efficient fashion.
\MethodName~uses frequency manipulation on the training images as a tool to control the size of the Gaussians and deals with low pass filters, which only keep low frequencies or large scale structures while discarding high frequencies containing texture patterns and noise components.
We experiment with mean, bi-lateral, Gaussian and Sobel filtering techniques in our approach and demonstrate that mean filtering is the most effective, shown in Section \ref{sec:ablation}.
A simple low pass filtering operation on an image can be described as:

\begin{equation}
	G[i,j] = \sum_{u=-k}^{k} \sum_{v=-k}^{k} H[u,v] F[i-u, j-v]
\end{equation}
where \( H \) and \( F \) represent the convolutional kernel and the input image respectively with \(G\) being the resulting image from the operation.
Importantly, we fix the stride to 1 for all our experiments and adjust the padding to ensure the spatial dimensions of the image don't change during any step of the image processing pipeline.
The values in the filtering kernel in the context of a convolution operation define the type of filtering that will happen to the image.
Gaussian kernels emphasize the central values in the filtering window of the image whereas a mean filter weighs them equally.
\subsection{Frequency Training Schedule}
\begin{figure}[!tbh]
    \centering
    \includegraphics[width=0.8\linewidth]{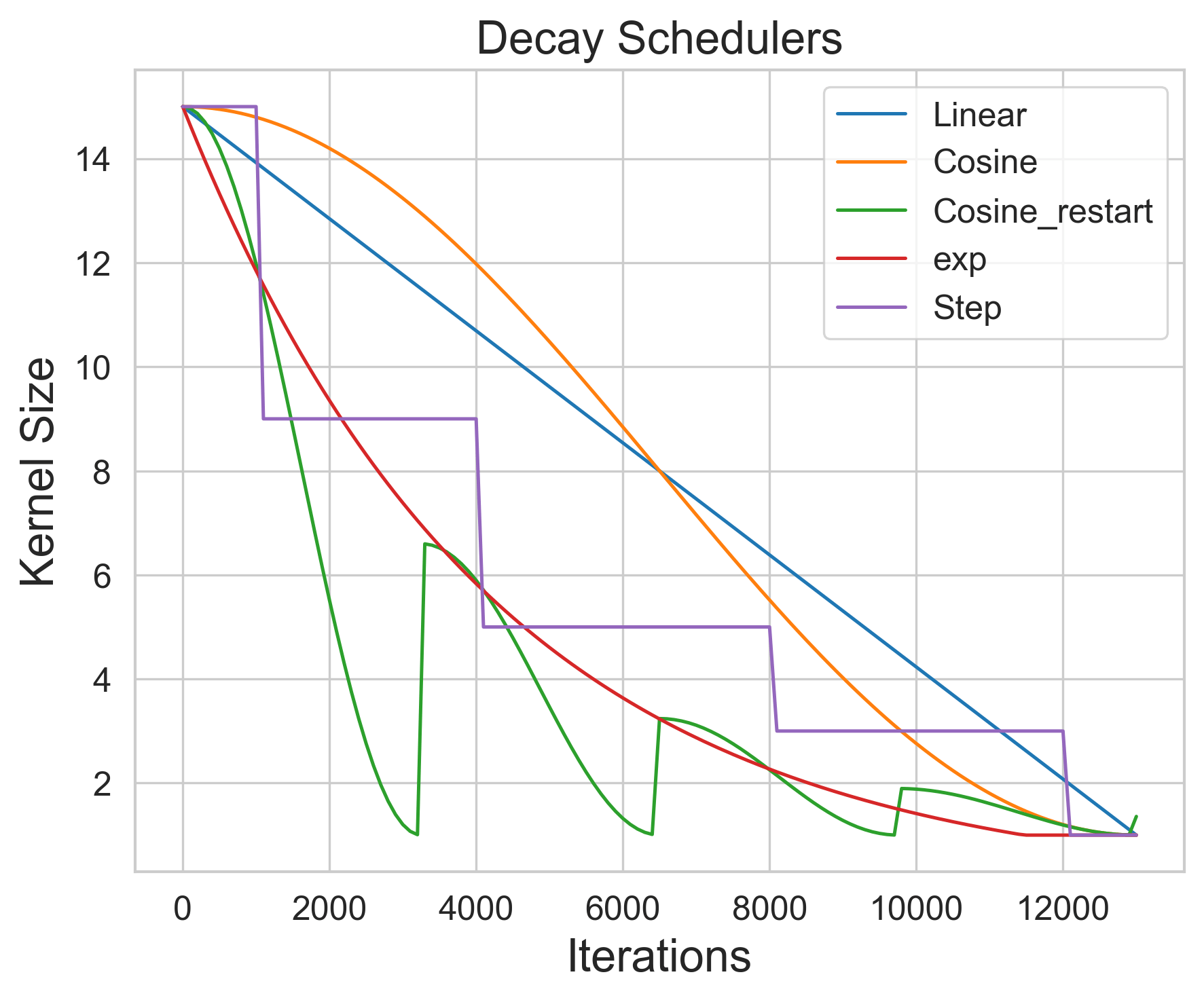}
    \caption{Visualization of different decay rates reported in our work. The vertical axis shows the current kernel size for the filtering algorithm for any given iteration. Note we stop all frequency modulation after 12,000 iterations and pass full sized images for the remaining 18,000 iterations.}
    \label{fig:decayCurves}
\end{figure}

Starting with very large coarse Gaussians and progressively adding high frequency details requires a schedule of frequency removal and addition.
We explore various mechanisms, including linear, cosine, cosine restarting, exponential and step functions, as shown in Figure \ref{fig:decayCurves}, which control the intensity of the image frequency modulation given the current training iteration as input.
For the step function we define a single value for a fixed iteration interval and define all intervals accordingly.
Based on ablation experiments, described in Section \ref{sec:ablation}, the step function to be most effective.

\subsection{Progressive Growing Pipeline} 
\MethodName~employs a progressive frequency control strategy with five distinct image and scene quality levels \(I_n\) and \(G_n\) respectively in range \(n=(1-5)\) as depicted in Fig.\ref{fig:teaser}.
We start by first obtaining a sparse scene reconstruction and initial 3D points using SfM on the training images \(I\) and this initial sparse representation acts as our \(G_0\) at \(n=0\) for initializing the Gaussians.
\begin{equation}
	\label{eq:gauss_l0}
	G_0 = SfM(I)
\end{equation}
Then for each subsequent level in our coarse-to-fine optimization process we use the last set of Gaussians \(G_{n-1}\) as the starting point and the normal 3DGS optimization loss is used to optimize \(G_{n-1}\) using the filtered images \(I\) in range \(n=(1-5)\).
\begin{equation}
	\label{eq:gauss_ln}
	G_n = 3DGS({G_{n-1}}, I_n)
\end{equation}
where the images \(I_n\) are obtained from the filtering function \(B(I, K_n, P_n)\). 
For each level the images get progressively sharper and more high frequency content is allowed to remain in the image. 
For the last level, \(n=5\),  we directly pass the original training images to the 3DGS optimizer.
\begin{equation}
	\label{eq:gauss_l4}
	G_5 = 3DGS({G_4}, I)
\end{equation}
For level-of-detail representation results, the model for any given level is saved at the last iteration before transitioning to the next level. The number of levels as well as the decay schedule can be easily changed for downstream applications.

	\begin{figure}[!tbh]
		\centering
		\includegraphics[width=0.8\linewidth]{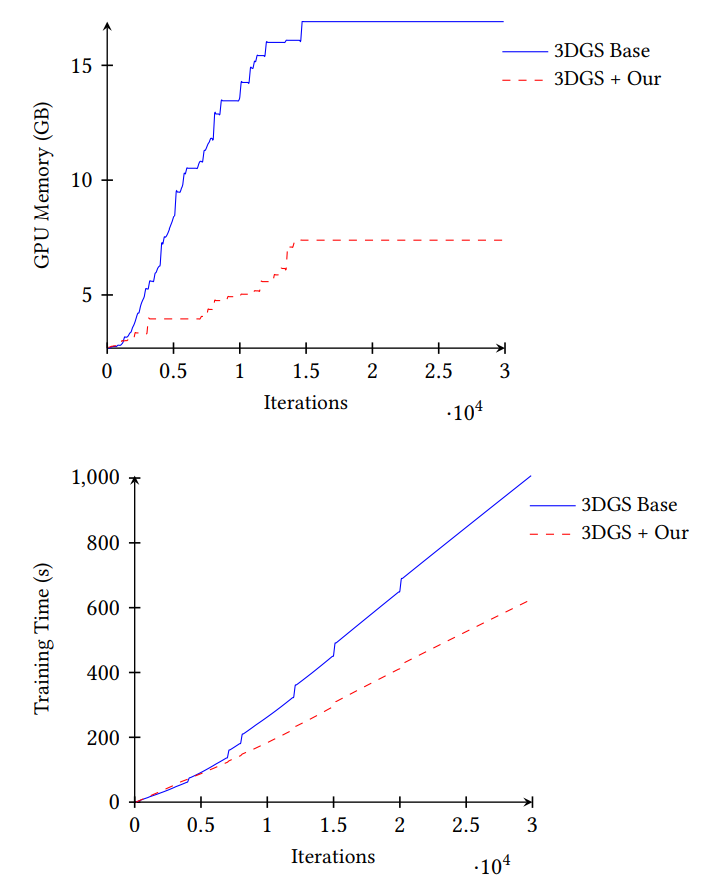}
	\caption{We reduce the GPU memory required during optimization due to a reduced Gaussian primitive load, which also reduces the total optimization time. These results are for 3DGS\cite{kerbl20233d} across the bicycle scenes and are represntive of all scenes tested.}
\label{fig:memtime}
	\end{figure}

	\begin{figure}[!tbh]
		\centering
		\includegraphics[width=0.8\linewidth]{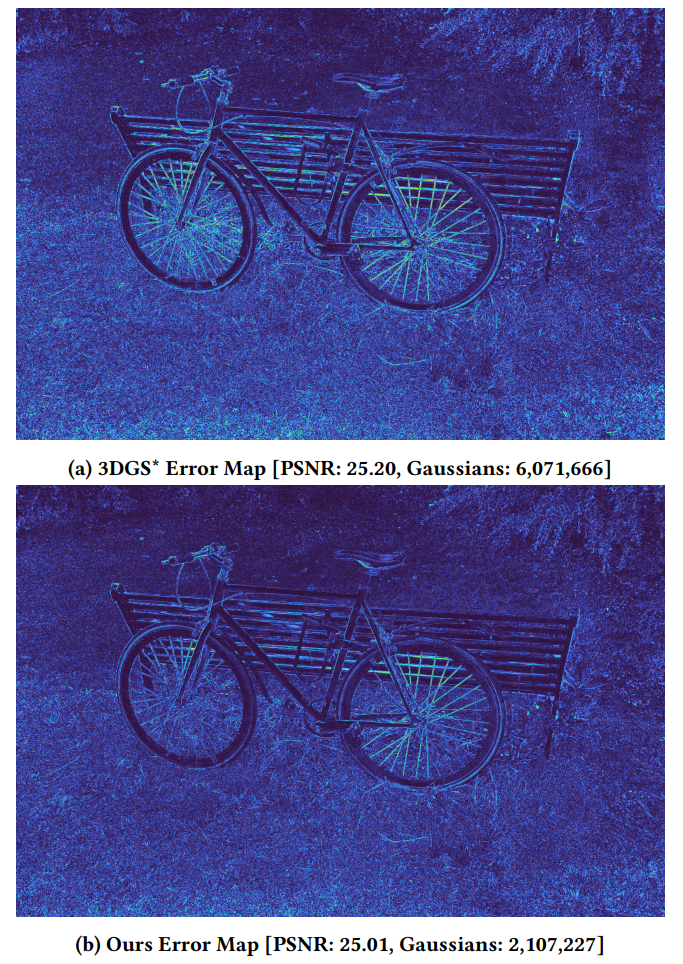}
	\caption{Pixel level comparison of error maps against the ground truth image. Our approach maintains similar quantitative scores while using fewer Gaussians and results in a decreased pixel wise error. Note the absence of noise like bright points in our results, due to the removal of tiny redundant Gaussians.}
		\label{fig:errorpixel}
	\end{figure}

\begin{table*}[t]
	\centering
	\caption{\MethodName~consistently lowers Gaussian count for all reported methods as well as improving the reconstruction quality for some. Note we also report the maximum Gaussian count for methods where it peaks before the end of optimization. No optimization parameters were changed for any method except for 3DGS+Ours$\ast$ as described in section \ref{sec:results}.}
	\resizebox{\textwidth}{!}{%
		\begin{tabular}{|c|c|c|c|c|c|c|c|c|c|c|c|c|c|c|c|c|c|c|c|c|c|}
			\hline
			\multirow{2}{*}{\textbf{Method}} & \multicolumn{7}{c|}{\textbf{Deep Blending}\cite{hedman2021baking}} & \multicolumn{7}{c|}{\textbf{Tanks \& Temples}\cite{Knapitsch2017}} & \multicolumn{7}{c|}{\textbf{Mip-NeRF360}\cite{barron2022mipnerf360}} \\
			\cline{2-22}
			& SSIM $\uparrow$ & PSNR $\uparrow$ & LPIPS $\downarrow$ & FPS $\uparrow$ & \begin{tabular}{@{}c@{}} Time \\ $(m:s)$\end{tabular} $\downarrow$ &  \begin{tabular}{@{}c@{}}\#G \\ ($10^6$)\end{tabular} $\downarrow$  & \begin{tabular}{@{}c@{}} Peak \\ \#G\end{tabular} $\downarrow$ 
			& SSIM $\uparrow$ & PSNR $\uparrow$ & LPIPS $\downarrow$ & FPS $\uparrow$ & \begin{tabular}{@{}c@{}} Time \\ $(m:s)$\end{tabular} $\downarrow$ & \begin{tabular}{@{}c@{}}\#G \\ ($10^6$)\end{tabular} $\downarrow$   & \begin{tabular}{@{}c@{}} Peak \\ \#G\end{tabular} $\downarrow$ 
			& SSIM $\uparrow$ & PSNR $\uparrow$ & LPIPS $\downarrow$ & FPS $\uparrow$ & \begin{tabular}{@{}c@{}} Time \\ $(m:s)$\end{tabular} $\downarrow$ & \begin{tabular}{@{}c@{}}\#G \\ ($10^6$)\end{tabular} $\downarrow$  & \begin{tabular}{@{}c@{}} Peak \\ \#G\end{tabular} $\downarrow$ \\
			
			\hline
			Compact-3DGS\cite{lee2023compact}        
			& 0.904 & 29.79 & \textbf{0.254} & 274 & 15:41       & 1.052 & 1.428 
			& \textbf{0.834} & \textbf{23.34} & \textbf{0.199} & 150 & 10:33 & 0.839 & 1.006 
			& \textbf{0.799} & \textbf{27.01} & \textbf{0.242} & 275 & 16:53 & 1.489 & 2.400  \\
			\rowcolor{gray!20}Compact-3DGS + Ours    
			& \textbf{0.904} & \textbf{29.81} & 0.258 & \textbf{348} & \textbf{13:46} & \textbf{0.831}\textcolor{blue}{\scriptsize~(-21\%)}  &  \textbf{1.130} 
			& 0.827 & 23.26 & 0.217 & \textbf{218} & \textbf{08:31} & \textbf{0.601}\textcolor{blue}{\scriptsize~(-28\%)} & \textbf{0.718} 
			& 0.792 & 26.96 & 0.262 & \textbf{357} & \textbf{13:37} &  \textbf{1.048}\textcolor{blue}{\scriptsize~(-30\%)} & \textbf{1.794} \\
			\hline
			
			Reduced-3DGS\cite{papantonakis2024reducing}        
			& 0.903 & 29.63 & \textbf{0.248} & 420 & 11:00& 0.925 & - 
			& \textbf{0.841} & 23.52 & \textbf{0.186} & 258 & 06:45 & 0.568 & - 
			& \textbf{0.810} & \textbf{27.29} & \textbf{0.229} & 408 & 11:55 & 1.460 & - \\
			\rowcolor{gray!20}Reduced-3DGS + Ours              
			& \textbf{0.904} & \textbf{29.76} & 0.251 & \textbf{558} & \textbf{09:53} & \textbf{0.697}\textcolor{blue}{\scriptsize~(-25\%)}& - 
			& 0.834 & \textbf{23.52} & 0.203 & \textbf{352} & \textbf{05:38} & \textbf{0.409}\textcolor{blue}{\scriptsize~(-28\%)} & - 
			& 0.801 & 27.14 & 0.250 & \textbf{560} &  \textbf{09:55}& \textbf{1.061}\textcolor{blue}{\scriptsize~(-27\%)} & - \\
			\hline
			
			Mini-Splatting\cite{fang2024mini}        
			& 0.904 & 29.88 & 0.256 & 839 & \textbf{10:00} & 0.352 & 0.378 
			& 0.832 & \textbf{23.22} & \textbf{0.202} & 602 & \textbf{07:10} & 0.203 & 0.217 
			& 0.820 & 27.25 & \textbf{0.219} & 699 & \textbf{11:01} & 0.496 & 0.611 \\
			\rowcolor{gray!20}Mini-Splatting + Ours  
			& \textbf{0.908} & \textbf{30.01} & \textbf{0.254} & \textbf{849} & 10:07 & \textbf{0.344}\textcolor{blue}{\scriptsize~(-2\%)} & \textbf{0.376} 
			& \textbf{0.834} & 23.17 & 0.204 & \textbf{614} & 07:13 & \textbf{0.197}\textcolor{blue}{\scriptsize~(-3\%)} & \textbf{0.211} 
			& \textbf{0.821} & \textbf{27.26} & \textbf{0.219} & \textbf{711} & 11:21 &  \textbf{0.488}\textcolor{blue}{\scriptsize~(-2\%)} &  \textbf{0.610} \\
			\hline
			Taming3DGS\cite{mallick2024taming} (budget)       
			& 0.899  & 29.75 & 0.273 & \textbf{1038} & \textbf{02:14} & 0.294 & - 
			& 0.832 & 23.66 & \textbf{0.213} & 411 & \textbf{02:19} & 0.319 & - 
			& \textbf{0.793}   & 27.21  & \textbf{0.262} & 523 & 14:36 & 0.666 & - \\
			\rowcolor{gray!20}Taming3DGS (budget) + Ours       
			& \textbf{0.905} & \textbf{30.00} & \textbf{0.269} & 1036 & 02:22& {0.294}\textcolor{blue}{\scriptsize~(0\%)} & - 
			& \textbf{0.834} & \textbf{23.90} & 0.215 & \textbf{431} & 02:17 &\textbf{ 0.318}\textcolor{blue}{\scriptsize~(0\%)} & - 
			& 0.788 & \textbf{27.27} & 0.274 & \textbf{593} & \textbf{14:11} & \textbf{0.5735}\textcolor{blue}{\scriptsize~(-14\%)} & - \\
			\hline
			Taming3DGS\cite{mallick2024taming} (big)       
			& 0.909  & 30.04 & 0.233 & 326 & 06:15 & 3.593 & - 
			& \textbf{0.859} & 24.14 & \textbf{0.160} & 107 & 05:04 &  2.290 & - 
			& \textbf{0.823} & \textbf{27.82} & \textbf{0.207} & 228 &   06:46& 3.251   & - \\		
			\rowcolor{gray!20} Taming3DGS (big) + Ours      
			& \textbf{0.911} & \textbf{30.13} & \textbf{0.237} & \textbf{431} & \textbf{04:57} & \textbf{2.212}\textcolor{blue}{\scriptsize~(-38\%)} & - 
			& 0.854 & \textbf{24.22} & 0.175 & \textbf{163} &  \textbf{03:38} & \textbf{1.333}\textcolor{blue}{\scriptsize~(-42\%)} & - 
			& 0.817 & 27.77 & 0.223 & \textbf{318} & \textbf{05:19} & \textbf{2.345}\textcolor{blue}{\scriptsize~(-28\%)} & - \\
			\hline
			EAGLES\cite{girish2024eagles}          
			& 0.91   & 29.86  & \textbf{0.25} & 247 & \textbf{11:22}& 1.19 & - 
			& \textbf{0.84} & 23.27  & \textbf{0.20} & 379 & \textbf{06:19} & 0.65 & - 
			& \textbf{0.81} & \textbf{27.23} & \textbf{0.24} & 274 & \textbf{09:40}&  1.33& - \\
			\rowcolor{gray!20}EAGLES + Ours         
			& \textbf{0.91} & \textbf{29.86}  & \textbf{0.25} & \textbf{304} & 11:36 & \textbf{0.87}\textcolor{blue}{\scriptsize~(-27\%)} & - 
			& 0.83 & \textbf{23.28} & 0.21 & \textbf{417} & 07:00 & \textbf{0.58}\textcolor{blue}{\scriptsize~(-11\%)} & - 
			& 0.79  & 26.90 & 0.27 & \textbf{329} & 13:21&  \textbf{0.90}\textcolor{blue}{\scriptsize~(-32\%)}& - \\
			\hline
			3DGS\cite{kerbl20233d}         
			& 0.900 & 29.54 & \textbf{0.247} & 293 & 11:34 & 2.795 & - 
			& \textbf{0.844} & \textbf{23.69} & \textbf{0.178} & 108 & 07:19& 1.825  & - 
			& \textbf{0.813} & \textbf{27.53} & \textbf{0.221} & 247 & 11:42& 3.157  & - \\
			\rowcolor{gray!20}3DGS + Ours$^{\ast}$
			& \textbf{0.902} & \textbf{29.56} & 0.255 & \textbf{611} & \textbf{10:26} & \textbf{0.953}\textcolor{blue}{\scriptsize~(-66\%)} & - 
			& 0.823 & 23.15 & 0.217 & \textbf{258} & \textbf{05:56} &  \textbf{0.658}\textcolor{blue}{\scriptsize~(-64\%)} & - 
			& 0.809 & 27.09 & 0.237 & \textbf{308} & \textbf{09:50} & \textbf{1.327}\textcolor{blue}{\scriptsize~(-58\%)} & - \\
			\hline
	\end{tabular}}
	\label{tab:main}
\end{table*}

\section{Results and Evaluation}
\label{sec:results}
\textbf{Datasets}: We report results on the MipNeRF360 \cite{barron2022mipnerf360}, DeepBlending \cite{hedman2021baking} and Tanks and Temples \cite{Knapitsch2017} datasets with a total of 13 scenes which include a good mix of indoor and outdoor unbounded scenes, as well as scenes with low-frequency textures.
%
%\textbf{Datasets}: Following common practice we use the MipNeRF360 \cite{barron2022mipnerf360}, DeepBlending \cite{hedman2021baking} and Tanks and Temples \cite{Knapitsch2017} datasets with a total of 13 scenes including a good mix of indoor and outdoor unbounded scenes.
%
%We use the train and truck scene from \cite{Knapitsch2017}, the drjohnson and playroom scene from \cite{hedman2021baking} and bicycle, bonsai, treehill, counter, kitchen, stump, flowers, garden and room scenes from \cite{barron2022mipnerf360}.

%
%
%
\noindent\textbf{Implementation Details}: All our ablations, graphs and results are reported as the average scores for all held-out test views averaged for all 13 scenes, with the exception of Figure \ref{fig:teaser} and Figure \ref{fig:errorpixel} which report the results averaged for all test views for the bicycle scene.
For testing the impact of our module on top of other 3DGS based methods, we first reproduce the software environment and reported results from the authors and then re-run the code with our changes applied using the authors provided training and evaluation scripts for fair comparison and evaluation.
We do not change any hyper-parameters or settings for any of the methods except for 3DGS \cite{kerbl20233d} where we change the \textit{scaling\_lr}$=0.01$, \textit{densification\_interval}$=500$ and \textit{densify\_grad\_treshold}$=0.0001$.
All experiments and comparison were undertaken using a NVidia H200 GPU.
We use the step decay function as shown in Figure \ref{fig:decayCurves} with a mean blur filter with an initial kernel size of 15 in all our experiments and stop frequency modulation after the first 12,000 iterations.
Further ablations of these parameters are presented in Section \ref{sec:ablation}.
Reconstructed scene models and code to reproduce results will be made publicly available.

\noindent\textbf{Metrics}: We use PSNR, SSIM and LPIPS for quantitative evaluation and analysis of the trained models and for reproducing relevant baselines.
We follow the established best practice of keeping every 8$^{th}$ frame for a hold-out test set and train all our experiments for 30,000 iterations and comply with the image resolution convention from \cite{kerbl20233d} for using full scale images for DeepBlending \cite{hedman2021baking} and Tanks and Temples \cite{Knapitsch2017} and using 2$^{nd}$ and 4$^{th}$ image resolution levels after Colmap processing for MipNeRF360 \cite{barron2022mipnerf360} indoor and outdoor scenes respectively.

\subsection{Progressive Growing Impact}
Our approach achieves excellent reduction in the number of Gaussians used to represent a scene while maintaining competitive visual quality or improving the visual quality.
\newline
\noindent\textbf{Reduced Gaussian Footprint}:We achieve a reduction in the number of Gaussians in all datasets for each of the methods as reported in Table \ref{tab:main}, which is impressive given that all listed methods aim to reduce the file size and Gaussian count and are already extremely optimized to this end. Averaged across all reported methods for the three datasets, there is a drop of -0.03dB in PSNR for an average drop of 26\% in the number of Gaussians. This represents a good trade-off with a - 0.13dB PSNR drop for a 27\% drop in \#G for Mip-NeRF360, a +0.09dB PSNR gain for a 26\% drop in \#G for Deep Blending and a - 0.05dB PSNR drop for a 25\% drop in \#G for Tanks\&Temples. The single largest reported PSNR drop in Table \ref{tab:main}. of 0.44dB equals a quality loss of 10\% while reducing \#G by 58\% and train time by 16\%.
\newline
\noindent When applied on 3DGS\cite{kerbl20233d}, our method reduces the Gaussian count by up-to 62\% as reported in Table \ref{tab:main}.
\newline
\noindent\textbf{Peak Gaussian Count}: For some methods, such as  MiniSplatting \cite{fang2024mini} and Compact3DGS \cite{lee2023compact}, the final Gaussian count is not a true representation of the GPU memory requirements as the Gaussian count peaks at various points during training.
\MethodName~also reduces the peak Gaussian load as well as the final Gaussian count for both MiniSplatting\cite{fang2024mini} and Compact3DGS\cite{lee2023compact} as reported in Table \ref{tab:main}.
\newline
\noindent\textbf{Reduced Optimization Time and GPU Memory}: A reduction in the number of Gaussians during the optimization process also leads to accumulated speed-ups in terms of optimization time as reported in Table \ref{tab:main} with an illustrative example visualized in the Figure \ref{fig:memtime}. For 3DGS\cite{kerbl20233d}, our method reduces the optimization time by 20\% and the training GPU memory requirements by up-to 40\%. Notably, the optimization time for EAGLES\cite{girish2024eagles} and Taming3DGS\cite{mallick2024taming}(budget) slightly increases, but this could potentially arise from other hyper-parameters in those methods being optimized for a different training regimes and having a high degree of coupling. Note, we do not change any parameters in any of these methods except for 3DGS. Still, the Gaussian footprint is reliably reduced. The GPU memory required for optimization also decreases proportionally to the reduction in the Gaussian count. An illustrative example of GPU memory reduction is highlighted in Figure \ref{fig:memtime}.

	\begin{figure*}[!tbh]
		\centering
		\includegraphics[width=\linewidth]{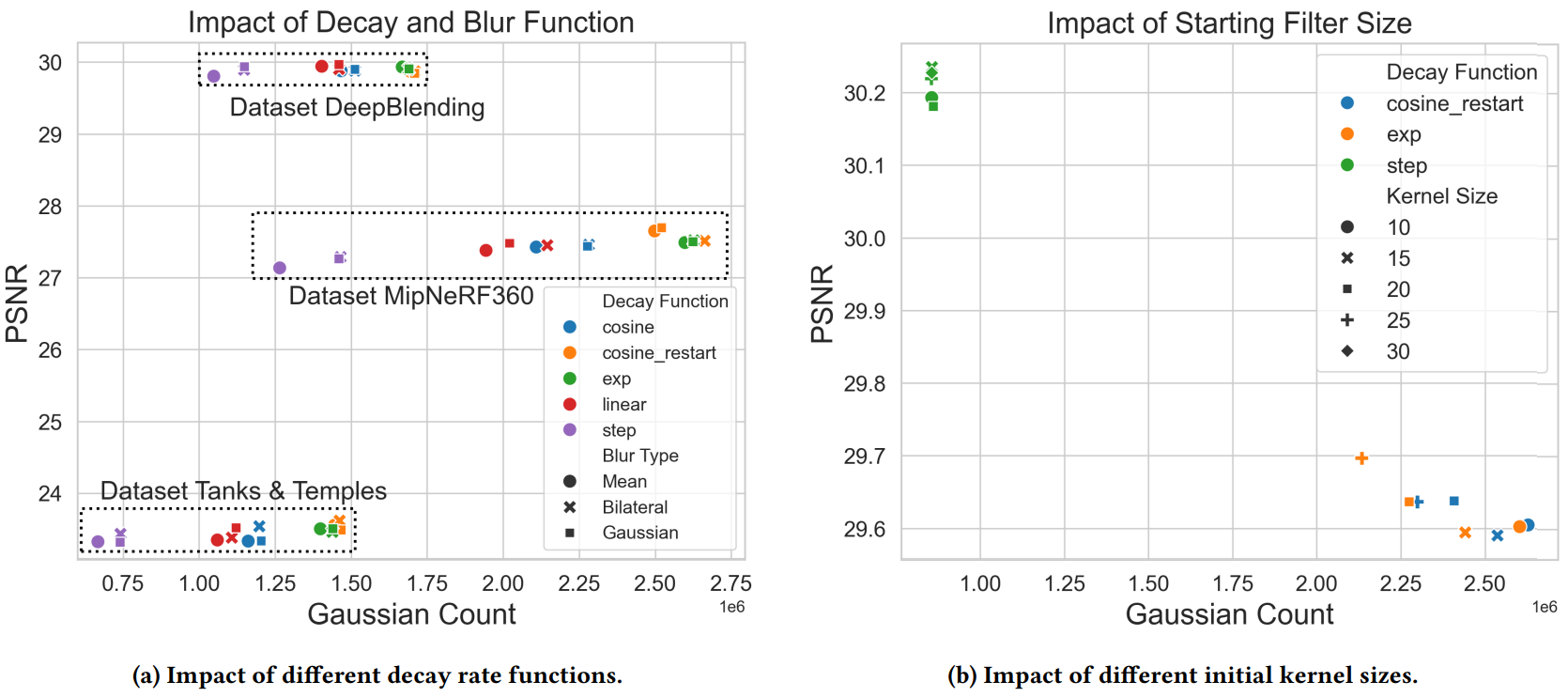}
			\caption{Ablation Experiments: (a) shows the impact of using different decay rate functions to control the size of the kernel used for image filtering. Combining step decay function with the mean filter performs best and is adopted across all our experiments. Note each point is the average score for that respective dataset. (b) shows the impact of using different initial kernel sizes with the mean filter. Larger starting kernel size reduces the Gaussian count but the quality plateaus.}
		\label{fig:ablation}
	\end{figure*}

\noindent\textbf{Level-of-Detail Representation}: \MethodName also produced different levels-of-details as a natural by-product of our progressive growing approach. We currently only store the last level-of-detail to save disk space. However, it allows a scene to be trained to the maximum level-of-detail before running out of available GPU memory. Figure. \ref{fig:memtime} and Figure \ref{fig:gaussianSize} indicate this might be an useful feature and intermediate representations from \MethodName might be more useful than ones obtained naively from 3DGS. We leave further exploration of this idea as future work.

\noindent\textbf{Performance on Scenes with Low Texture:} \MethodName~ gives the highest gains in visual quality on DeepBlending where all scenes include large untextured walls and ceilings with the PSNR improving on all seven reported methods. There is a gain of +0.09dB in PSNR for a 26\% drop in \#G for Deep Blending and we leave further investigation of this result in context of the SfM induced bias to future work.
    \begin{table}[!tbh]
	\centering
	\caption{
		Compares our progressive filtering approach to the progressive downsampling approach proposed by \cite{girish2024eagles}. Both are applied on 3DGS\cite{kerbl20233d} in isolation without any additional changes. We use the default image downsizing based approach with starting resize threshold of 0.3 and duration of 0.7 for \cite{girish2024eagles}. Our progressive filtering uses the step decay function with mean blur filter and an initial kernel size of 15, which is consistent with the results reported in Table. \ref{tab:main}. Results indicate that both approaches have a distinct impact on different datasets.}
	\resizebox{\linewidth}{!}{%
			\begin{tabular}{|c|c|c|c|c|c|}
				\hline
				 Dataset &  Method &  SSIM $\uparrow$ & PSNR $\uparrow$ & LPIPS $\downarrow$ & \begin{tabular}{@{}c@{}}\#G \\ ($10^6$)\end{tabular} $\downarrow$ \\
				\hline
				\multirow{2}{*}{Mip-NeRF360}  
				& EAGLES\cite{girish2024eagles}-Downsizing & \textbf{0.810} & \textbf{27.44} & \textbf{0.232} & 2.412 \\
				~ & Ours-Filtering 							  & {0.808} & {27.37} & {0.238} & \textbf{2.195}  \\
				\hline
				\multirow{2}{*}{Tanks \& Temples}  
				& EAGLES\cite{girish2024eagles}-Downsizing & 0.843 & 23.59 & {0.191} & \textbf{1.090} \\
				& Ours-Filtering                               & \textbf{0.843} & \textbf{23.71} & \textbf{0.191} & 1.229  \\
				\hline
				\multirow{2}{*}{Deep Blending}  
				& EAGLES\cite{girish2024eagles}-Downsizing & 0.904 & 29.55 & 0.247 & 2.325 \\
				& Ours-Filtering 							  & \textbf{0.906} & \textbf{29.68} & \textbf{0.246}  & \textbf{2.118}\\
				\hline
			\end{tabular}
		}
		\label{tab:eagle}
	\end{table}

\subsection{Efficient use of Gaussian Primitives}
For 3DGS \cite{kerbl20233d} our approach reduces the Gaussian primitive count by up-to 62\% but it also favours more efficient use of any given Gaussian budget, meaning it can derive more quality and a better reconstruction given the same number of Gaussian primitives. This explains the simultaneous increase of +0.09dB in Deep Blending dataset accompanied by a 26\% drop in number of Gaussians used.
The ADC in 3DGS generally adds redundant and even wasteful Gaussians in some parts of the scene driving up the compute requirements with no gain in quality \cite{niedermayr2023compressed, lee2023compact, papantonakis2024reducing}.
Our method reduces the wasteful and redundant use of Gaussians as well and drives the quality higher as shown by the lack of bright noise pattern seen in Figure \ref{fig:errorpixel}.
	\begin{figure*}[]
		\centering
		\includegraphics[width=\linewidth]{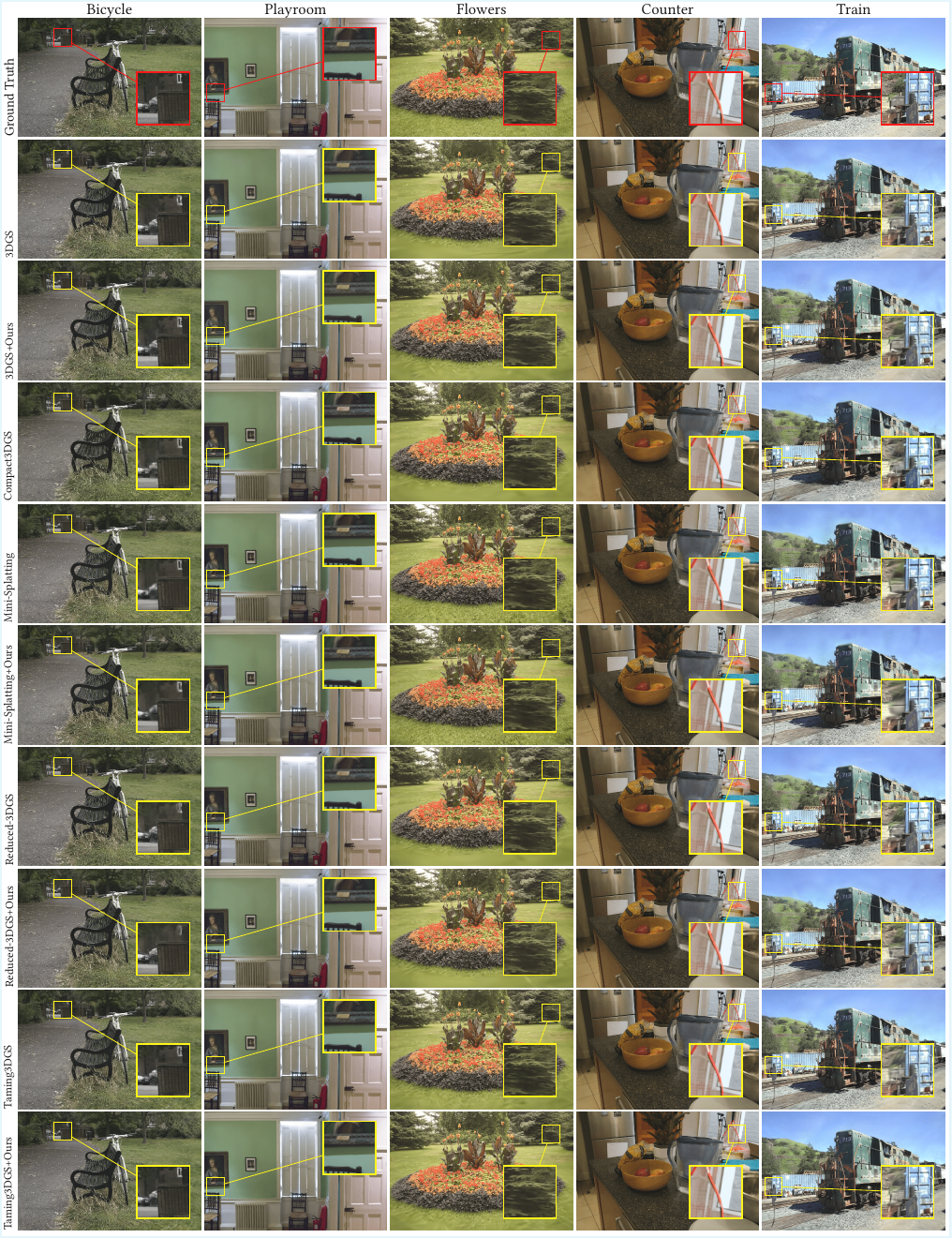}
		\caption{Qualitative results of \MethodName~when integrated into different 3DGS-based methods with zoomed in cut-outs.}
		\label{fig:fig1}
	\end{figure*}

	\begin{figure*}[!tbh]
		\centering
		\includegraphics[width=\linewidth]{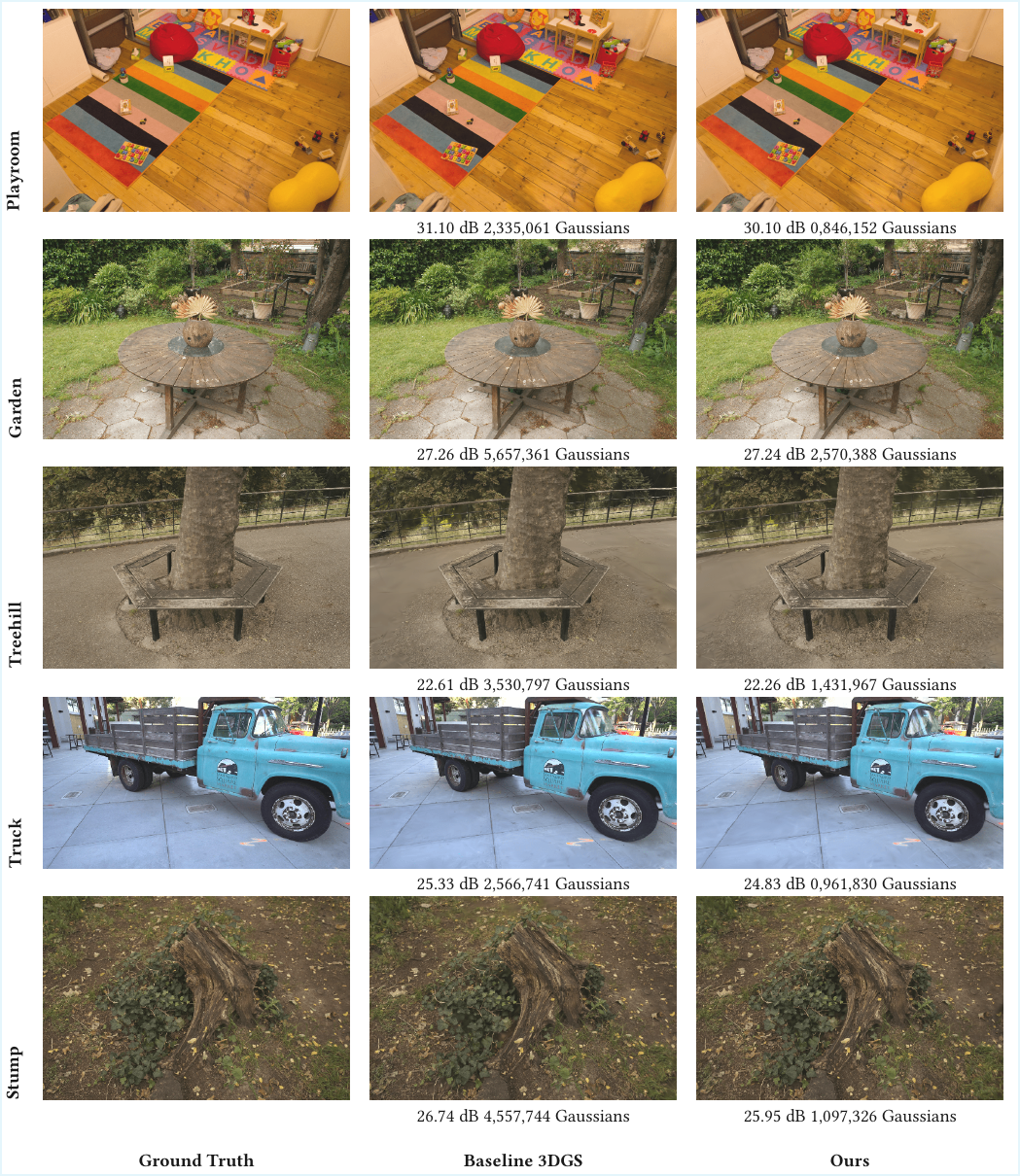}
		\caption{Comparison of our results with 3DGS and respective ground truths. Best viewed zoomed in.}
		\label{fig:fig2}
	\end{figure*}

\subsection{Ablations}
\label{sec:ablation}
\noindent\textbf{Modulation Schedule}: We experiment with various frequency modulation schedules in Figure \ref{fig:decayCurves}. The results indicate that the step schedule function works better than linear, exponential, or cosine variants across all scenes in the datasets by a significant margin. This is explained by the fact that it allows the ADC to optimize a fixed Gaussian budget rather than optimizing and simultaneously handling new insertions.
\newline
\noindent\textbf{Blur Filter}: We also test various types of image filtering algorithms including Gaussian, bilinear and mean filtering to determine the best approach. 
The results, shown in Figure \ref{fig:decayCurves}, indicate that the mean filter performs well on most type of scenes. 
Surprisingly Gaussian and bilateral filtering perform poorly, possibly because they give more weight to the pixel at the center of the kernel or to edge information in the images.
\newline
\noindent\textbf{Kernel Size}: Holding the step function constant and using a mean filter, we perform experiments to evaluate the impact of the image filtering intensity on the final reconstruction quality.
Results, shown in Figure \ref{fig:ablation}, indicate that starting from a larger filter improves performance but the trend peaks quickly. We adopt the step decay function with an initial kernel size of 15 in all our experiments for the first 12\_000 iterations.
\newline
\noindent\textbf{Improvement over EAGLES:}
EAGLES focuses on storage size and render and train time, starts from 3DGS and uses: (i) MLP based quantization (SH color, opacity and rotation attributes); (ii) influence based Gaussian pruning; (iii) progressive resolution training.
\MethodName~ only uses progressive frequency modulation and 3DGS+\MethodName~ when compared to the full highly optimized EAGLES pipeline achieves better average Gaussian count by 6.3\%, better average Train Time by 4.2\% and better average FPS by 43\% while there is a minor average PSNR loss of 0.18 dB.
\newline
\noindent\textbf{Full-EAGLES with Our Frequency Modulation:} 
\noindent `EAGLES + Ours' in Tab. \ref{tab:main} shows that our frequency modulation when plugged into the full EAGLES pipeline as a replacement for its progressive resolution downsampling improves average Gaussian count by 23\% with an average PSNR drop of 0.1 dB.
\noindent Slight loss in quality could be due to EAGLES hyperparameters being optimized for its own resolution modulation and we do not change any of them to ensure a fair and clean comparison.

\noindent\textbf{Core Novelty comparison with EAGLES:} 
{EAGLES progressive training module gradually increases image resolution} whereas {\MethodName~ only modulates image} {frequency} and when both are applied to 3DGS in isolation our frequency modulation shows a distinct and different impact from resolution downsampling. Our approach reduces Gaussian footprint for two of the three datasets as well as slightly improving the reconstruction quality for two datasets.
%
%
%
%
% Conclusions 
%
\section{Conclusions}
\noindent\textbf{Summary}: We have presented \MethodName, a simple yet highly effective framework that minimizes the number of Gaussian primitives in 3DGS-based pipelines.
We demonstrated an average reduction of 62\% in Gaussian count, a 40\% reduction in the training GPU memory requirements and a 20\% reduction in optimization time without sacrificing the visual quality compared to 3DGS \cite{kerbl20233d}.
We have conducted an extensive quantitative and qualitative evaluation that has shown \MethodName~can be seamlessly integrated into a wide range of current 3DGS-based approaches resulting in a positive impact on primitive count in all cases. 
The contribution of \MethodName~enables 3DGS-based pipelines and scene reconstructions to be more usable on a range of consumer-grade devices due to reduced primitives and therefore storage, transmission and utilization requirements making the technology more widely accessible. 

\noindent\textbf{Limitations}: Currently \MethodName~is agnostic to the image content and applies filtering on the entire training set uniformly, but it fails to take into account structural boundaries and large constant object which could be better represented by singular Gaussians with the right pixel level emphasis. 
A content-aware extension will be considered in the future.

\noindent\textbf{Future Work} will focus on using edge and depth aware image filtering operations and investigating the impact of using projected Gaussian centroid density in the image space to selectively guide attention to areas with lower visual variation which remain poorly densified by the ADC. 
We will also explore the extension of \MethodName~to dynamic scenes exploiting the level-of-detail representation.

\bibliographystyle{ACM-Reference-Format}
\bibliography{main}

\end{document}